\documentstyle[twocolumn,prl,aps]{revtex}

\begin{document}

\twocolumn[\hsize\textwidth\columnwidth\hsize\csname @twocolumnfalse\endcsname
\draft 
\preprint{} 
\title{Quantum Hall effect at weak magnetic field: New float-up picture}
\author{D.N. Sheng$^1$, Z.Y. Weng$^1$, and X.G. Wen$^2$} 
\address{$^1$Texas Center for Superconductivity, 
University of Houston, Houston, TX 77204-5506}  
\address{$^2$Department of Physics, Massachusetts Institute of Technology, 
Cambridge, MA 02139} 
\maketitle
\date{today}
\begin{abstract} 

The fate of integer quantum Hall effect (IQHE) at weak magnetic field is studied
numerically in the presence of {\it correlated} disorders. We find a systematic 
{\it float-up} and {\it merging} picture for extended levels on the low-energy 
side which results in direct transitions from higher-plateau IQHE states to 
the insulator. Such direct transitions are controlled by a quantum critical 
point with a {\it universal} scaling form of conductance. The phase diagram 
is in good agreement with recent experiments. The issue of continuum vs. 
lattice model is also discussed. 

\end{abstract}
]

How extended levels evolve with disorders and the magnetic field $B$ is
central to our understanding of the IQHE. Earlier on, Khmel'nitzkii\cite{khm}
and Laughlin\cite{laughlin} had argued that extended levels should
continuously float up towards higher energy with reducing $B$. And the
assumption that extended levels never merge has led to a select rule of the
global phase diagram\cite{klz} for IQHE, in which a direct
transition from a higher-plateau ($\nu >1$) state to the insulator is
prohibited. But direct transitions have been observed in many
recent experiments\cite{hilke,krav,song}, which have renewed the theoretical
interest to reexamine the float-up picture in IQHE systems.

Previous numerical studies in the tight-binding model (TBM) with 
{\it white-noise} disorders have indicated\cite{xie,dns,others,phased} the 
existence of direct transitions from higher IQHE plateau states to the 
insulator. 
But the detailed analysis\cite{dns} has also revealed that the lattice effect 
plays a central role there: such direct transitions first happen near the band 
center due to the presence of extended levels carrying negative topological 
Chern number (a peculiar lattice effect) which form a high-energy IQHE-insulator 
boundary and start to ``float-down'' towards the low-energy regime with 
increasing disorder or reducing $B$. During the float-down process, the boundary 
keeps merging with lower extended levels such that the plateaus disappear in a 
one-by-one fashion.

The key issue is if such a float-down picture, due to the lattice 
effect, is the unique explanation for direct transitions or there exists a 
different kind of direct transition free of the lattice effect. Recall that in
a continuum model, there does not exist a high IQHE-insulator boundary as the 
band center is essentially located at infinite energy. In this case a 
levitation of extended levels by disorders is generally expected as 
discussed\cite {haldane,fogler} perturbatively. Since it has been generally 
believed that the experimental situation should be physically described by the 
continuum model due to the weakness of magnetic fields compared to the 
bandwidth and low density of charge carriers, it becomes especially 
interesting whether the float-up of extended levels alone can also 
lead to a direct transition in the low-energy regime.

A float-up of the lowest extended level actually has been seen in the
numerical calculation\cite{phased} based on TBM but its journey has
quickly ended by merging into the float-down IQHE-insulator boundary from
the band center. In order to study how such a float-up feature near the band 
edge evolves, one has to somehow  ``delay'' the floating-down process of the 
high IQHE-insulator boundary. Note that the inter-Landau-level-mixing caused by
uncorrelated (white-noise) disorders happens more strongly near the band center, 
which may {\it enhance} the tendency for extended levels near the band center 
to move down. So one can try to ``smooth'' the lattice effect by introducing 
short-range {\it correlations} among disorders. As a result to be shown below,
a float-up process will then become a dominant effect for those {\it 
lower} extended levels as the float-down of the high IQHE-insulator boundary is
significantly slowed, in contrast to the case in the
white-noise limit\cite{xie,dns,others} at similar weak magnetic fields.

In this Letter, we present a systematic floating-up and {\it merging}
pattern revealed for extended levels near the band edge. Specifically, the
lowest extended level starts to float upward at stronger disorder or weaker $%
B$ and eventually emerges into the second lowest extended level to form a 
{\it new} IQHE-insulator boundary on the low-energy side, leading to a $%
\nu=2\rightarrow 0$ direct transition, while the aforementioned  
{\it upper} IQHE-insulator boundary still remains at high energy. And such a
lower IQHE-insulator boundary keeps moving up to merge with higher-energy
extended levels to result in $3-0$, $4-0$, ... direct transitions with
increasing disorders or reducing $B$. 
The phase diagram is in good agreement with the experiments. Furthermore, 
direct transitions to the insulator at the lower boundary are found to be 
consistent with a quantum critical point picture, and in particular the 
conductance as a function of $ B-B_c$ ($B_c$ denotes the critical 
magnetic field) is of the {\it universal} form for $1-0$, $2-0$, ... up 
to $6-0$ transitions within the numerical resolutions.  

The TBM model $H=-\sum_{<ij>}e^{ia_{ij}}c_i^{+}c_j+H.c.+
\sum_iw_ic_i^{+}c_i$, with the magnetic flux per plaquette $\phi=\sum_{\Box
}a_{ij}=2\pi m/M $ ($m$ and $M$ are integers). And we define $B=m/M$. The 
correlated disorder $w_i$ is generated by
$w_i= W/\pi  \sum_j f_j e^{-|{\bf R}_{i}-{\bf R}_{j}|^2/\lambda _0^2}$.
Here ${\bf R}_i$ denotes the spatial position of site $i$. $W$ and $\lambda
_0 $ are the  strength and correlation length scale of
disorders, respectively. $f_i $ is a random number distributing uniformly
between (-1,1).

To illustrate how extended levels near the band edge evolve with the
disorder strength $W$, the density of states carrying nonzero Chern number ($%
\rho_{ext}$) as a function of the Landau level filling number ($n_{L}$) is
plotted in Fig. 1 (a) and (b) at $B=1/64$, $\lambda _0=1$, and the sample 
size $32\times 64$. The sample-size-independent peaks in the figures denote the 
positions of
extended levels while the non-peak part of $\rho_{ext}$ should scale to zero
presumably in the thermodynamic limit\cite{dns,bhatt}. In Fig. 1(a), as $W$
is increased from $0.7$ to $1.4$, extended level positions (marked by
diamonds) all start to {\it float up} slightly from the filling number $%
n_{L}= \nu+0.5$ ($\nu=0, 1,...$). But the lowest extended level distinctly
moves faster and eventually {\it merges} into the second lowest extended
level to form a new boundary extended level [see $W=1.45$ and $1.55$ cases
in Fig. 1(b)], while the higher extended levels still remain roughly equally
spaced (note that in this strong disorder case, the merged peak of $\rho_{ext}$
becomes less significant but later we will discuss a more reliable way to
identify the extended levels). It corresponds to the {\it collapse} of the
mobility gap separating the lowest two neighboring extended levels and thus
the {\it destruction} of the $\nu=1$ IQHE plateau in between. By further
increasing $W$, we see that such a newly merged extended level boundary
continuously floats up and merges into the third extended level, and so on
and so forth. In this way, the IQHE plateaus disappear on the low $n_L$
side also in a one by one fashion. We emphasize that higher extended
levels originally at $n_{L}=\nu+0.5$ ($\nu \geq 1$) never move passing the
$n_{L}=\nu +1 $  before merging with the lower boundary [Fig. 1(b)] such that 
each IQHE plateau does not really float away before its destruction. 

Such a float-up and merging picture persists into very weak magnetic 
fields (from $B=1/64$ to $B=1/2304$). The  disorder strength 
$W_c$ at which the merged lowest two extended levels pass $n_L=2$, 
resulting in a $2-0$ transition, is shown in the inset of Fig. 1(c) as a 
function of $B$. We see that $W_c$ monotonically decreases with $B$ and is 
extrapolated to zero in zero $B$ limit in a fashion of $B^{1/2}$.

By following the trace of extended levels in the $n-B$ plane ($n$ is the
on-site electron density) at fixed $W=1.4$ and $\lambda_0=1$, we determine a
phase diagram in Fig. 1(c), where the filled circles represent the lower
IQHE-insulator boundary, while the open circles are the positions of various
extended levels between plateaus which merge into the boundary at weaker $B$.
This phase diagram is very similar to the recent experimental phase diagram
[Fig. 2 in Ref.\onlinecite{hilke}] as well as the earlier one obtained in
Ref.\onlinecite{krav,song}.  The Hall conductance calculation confirms that
$\sigma_{xy}$ indeed saturates to $\nu e^2/h$ in $\nu$-th plateau region
while it approches zero on insulating side. Both critical conductances 
$\sigma_{xxc}$ and $\sigma_{xyc}$ at $\nu\rightarrow 0$ transition are close 
to $\nu e^2/2h$ in accordance with experiments \cite{song}. 

Apparently the above results critically depend on how reliably one can
identify the positions of extended levels using finite-size calculations.
Let us focus on the merged extended levels as the lower IQHE-insulator
boundary shown in Fig. 1(c). By fixing $n_L= 2$, marked by the arrow 
$C$ in Fig. 1(c), we calculated the longitudinal conductance $\sigma_{xx}$ 
with $B$ changing continuously at fixed $W=1.4$. We found a peak in 
$\sigma_{xx}$ at $B_c=1/70$ as a $2-0$ transition. 
In Fig. 2(a), $\sigma_{xx}$ as a function of $B$ with $n_L=1,2, ... $, 
and $6$ at $W=1.4$ are shown for sample width $L=96$ (the stripe sample 
with $L_x=L$ and $L_y \sim 10^6$ is considered using transfer matrix 
method\cite{land}). The $L$-independent peak positions should correspond
to $\nu=1\rightarrow 0$, $2\rightarrow 0$, ..., and $6\rightarrow 0$ 
transitions in Fig. 1(c). Remarkably, all these data can be
collapsed onto a universal curve 
\begin{equation}  \label{2}
\sigma_{xx}/\sigma_c =2\exp(s)/(1+\exp(2s)),
\end{equation}
as shown in Fig. 2(b) if one defines a variable $s=c_{\nu}(L)(B-B_c)/B_c$. 
Here the
parameter $1/c_{\nu}(L)$ represents the relative width of $\nu\rightarrow 0$
transition at a finite $L$. Furthermore, by collapsing the data at different 
$L$'s, we find a scaling curve for $\nu=2\rightarrow 0$ transition in Fig.
3(a), with 
\begin{equation}  \label{1}
c(L) \propto L^{1/x}
\end{equation}
from $L=32$ to $L=160$. The correlation length exponent is identified to be $%
x=4.6\pm 0.5$ in the inset of Fig. 3(a), about doubled from $x=2.3$
for the $\nu=1\rightarrow 0$ transition which has been similarly determined
at $n_L=1$. The exponents for $3-0$, ..., $6-0$ transitions seem further
increased but are more difficult to determine with a similar accuracy for 
larger sample sizes are needed. It is noted that the scaling form (1) is 
not limited to $n_{L}=$ integer. For example, the data at $\nu=2 
\rightarrow 0$ transition with $n_{L}=2.2$ can be also collapsed onto the 
same curve. Furthermore, Eq.(1) still holds as we change $\lambda_0$ 
from $1$ to $2$ and $3$. Details will be presented elsewhere.

Based on (1) and (2), we conclude that the $\nu=2\rightarrow 0$ transition
corresponds to a quantum critical point with measure zero in the $%
L\rightarrow \infty$ limit. We note that the same scaling form has been
previously obtained\cite{mit} for the $1-0$ transition, where $%
\rho_{xx}=exp(-s)\frac h {e^2}$ and $\rho_{xy}= \frac h {e^2}$ leading to (1).
Identifying such a simple scaling relation for $\nu=1, 2\rightarrow 0$ as
well as higher plateaus to insulator transitions may be the most striking
evidence for a single quantum critical point at each transition. The standard 
scaling method
can be also applied to the $\nu=2 \rightarrow 0$ transition to {\it %
independently} verify the one parameter scaling law\cite{huk}. As shown in
Fig. 3(b), by collapsing the same  data  as $%
\sigma_{xx}(L)/\sigma_c=f(\xi/L)$ by a correlation length $\xi$, we find 
$\xi \propto |(B-B_c)/B_c|^{-x}$ with $x=4.5\pm 0.5$ [the inset of 
Fig. 3(b)] in agreement with the above result. 

We have also checked the case right before the lowest two extended levels merge 
together. By scanning $B$ at a fixed $n_L=1.8$ nearby the scan 
$C$ shown in Fig. 1(c), $\sigma_{xx}$ exhibits two distinct peaks at $B_{c1}=
0.0151$ and $B_{c2}=0.0169$ with $L=128$ [see the middle inset of
Fig. 4(a)]. The main panel of Fig. 4(a) shows the finite-size scaling curve
of $\sigma_{xx}/\sigma_c$ as a function of $\xi/L$ obtained by collapsing
the data of different sample sizes at $B<B_{c1}$. The right inset indicates
that $\xi$ diverges at $B_{c1}$ with an exponent $x=2.4$ which is
essentially the same as the standard one for the $1-0$ transition. Similar
finite-size scaling curve has been also obtained for the branch at $B>B_{c2}$%
, corresponding to the $2-1$ transition. It is noted that the above analysis
resembles the study\cite{zwang} of the spin unresolved case at strong
magnetic field where a small spin-orbit coupling is used to lift the spin
degeneracy to create two separated but very close quantum critical points.
Similar to the latter case, if one ``mistakenly'' treats the present case as
a {\it single} critical point at $B_m$, a middle point between $B_{c1}$ 
and $ B_{c2}$, and proceeds with a finite-size scaling analysis, then one
gets Fig. 4(b) where the quality of data collapsing becomes markedly worse
and in particular $\xi$ shows a saturation trend approaching $B_m$, contrary
to the assumption of a critical point at $B_m$. Thus, the two separated
extended levels with $|B_{c1}-B_{c2}|/B_m\sim 0.11$ is not mistakable as a 
single critical point in our numerical analysis. Finally, we make a remark that 
even if the 2-0 transition that we observed is actually two transitions with 
very small $|B_{c1}-B_{c2}|$ indistinguishable numerically, the nice scaling
properties [(1) and (2)] that we found still indicate that there is a new 
quantum critical point. In
this case, the splitting $|B_{c1}-B_{c2}|\neq 0$, if it exists, should be 
caused by a relevant operator of such a new critical point. 

To summarize, we have identified for the first time a new float-up and merging
pattern for extended levels near the band edge at weak $B$ (down to 
$B=1/2304$ where there are $1252$ Landau levels between the band edge and center).  The corresponding 
phase diagram with direct transitions is in excellent agreement with the 
experiments where the essential features can be explained by the 
narrowing and destruction of each IQHE plateau due to the sequential merging of 
neighboring extended levels as the mobility gap in between collapses.

{\bf Acknowledgments} - D.N.S. would like to acknowledge helpful
discussions with R. N. Bhatt, F.D.M. Haldane, and M. Hilke.  D.N.S. and
Z.Y.W. are supported by the State of Texas through ARP Grant No. 3652707 and 
TCSUH. X.G.W. is supported by NSF Grant No. DMR--97--14198 and  NSF-MRSEC 
Grant No. DMR--98--08941.

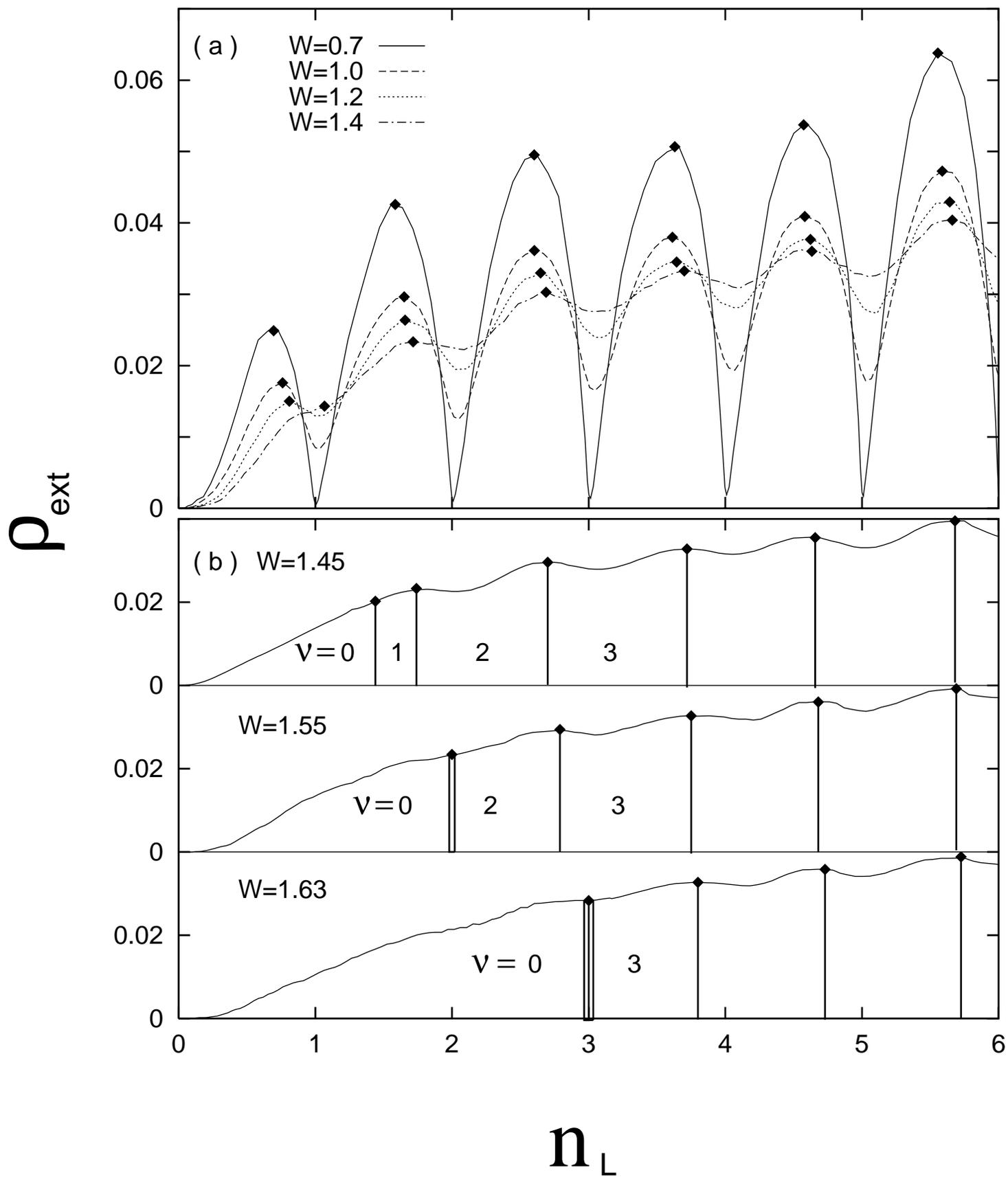
Fig. 1. (a) The postions of extended levels determined by the peaks 
(filled $\diamond$) of the density of states with nonzero Chern number (see
text). (b) Float-up and merging pattern of extended levels with the 
increase of the disorder strength $W$. (c) The numerical phase diagram in 
electron density - magnetic field ($n$-$B$) plane at $\lambda_0=1$.
The inset: The critical disorder $W_c$ for $\nu=2 \rightarrow 0$ transition 
at $n_L=2$ as a function of B.  
                                       
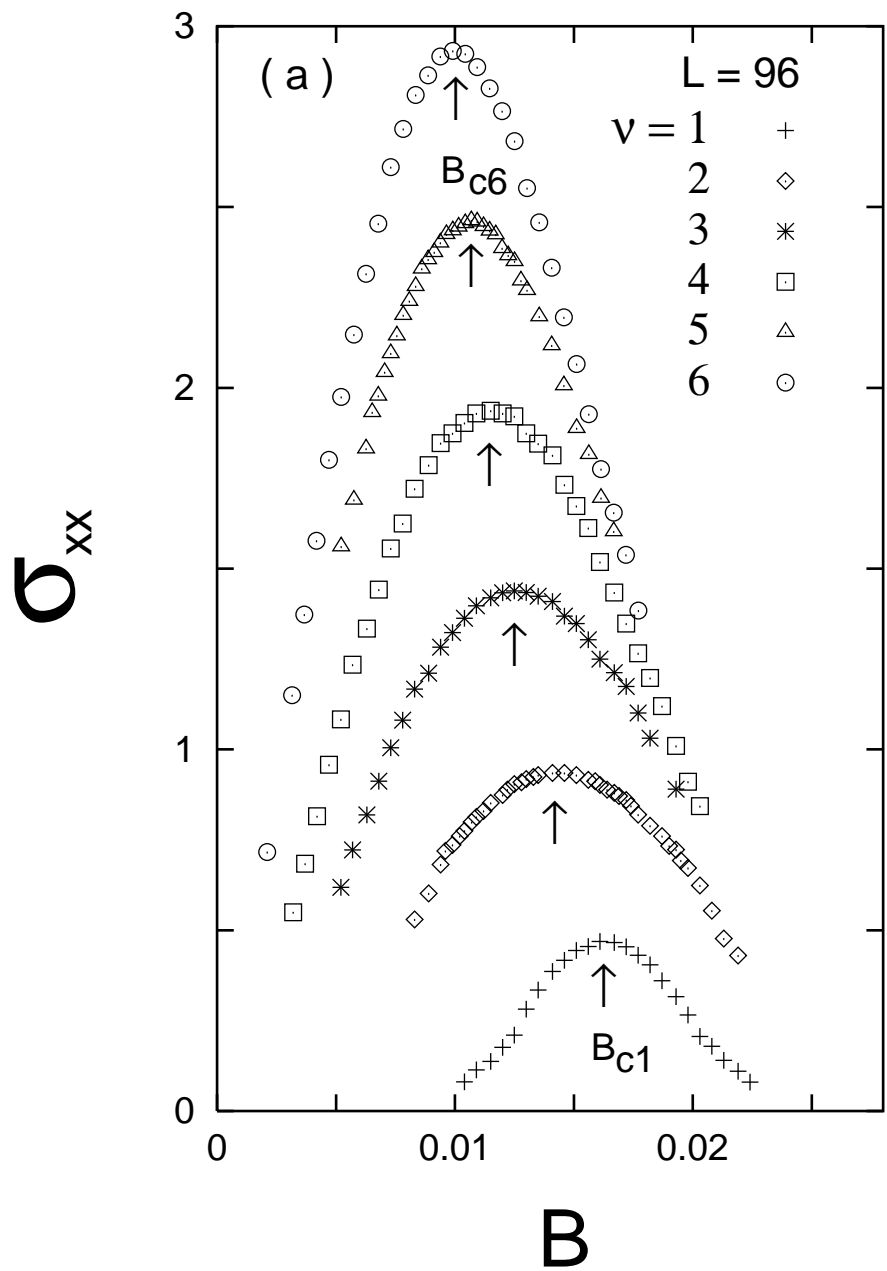
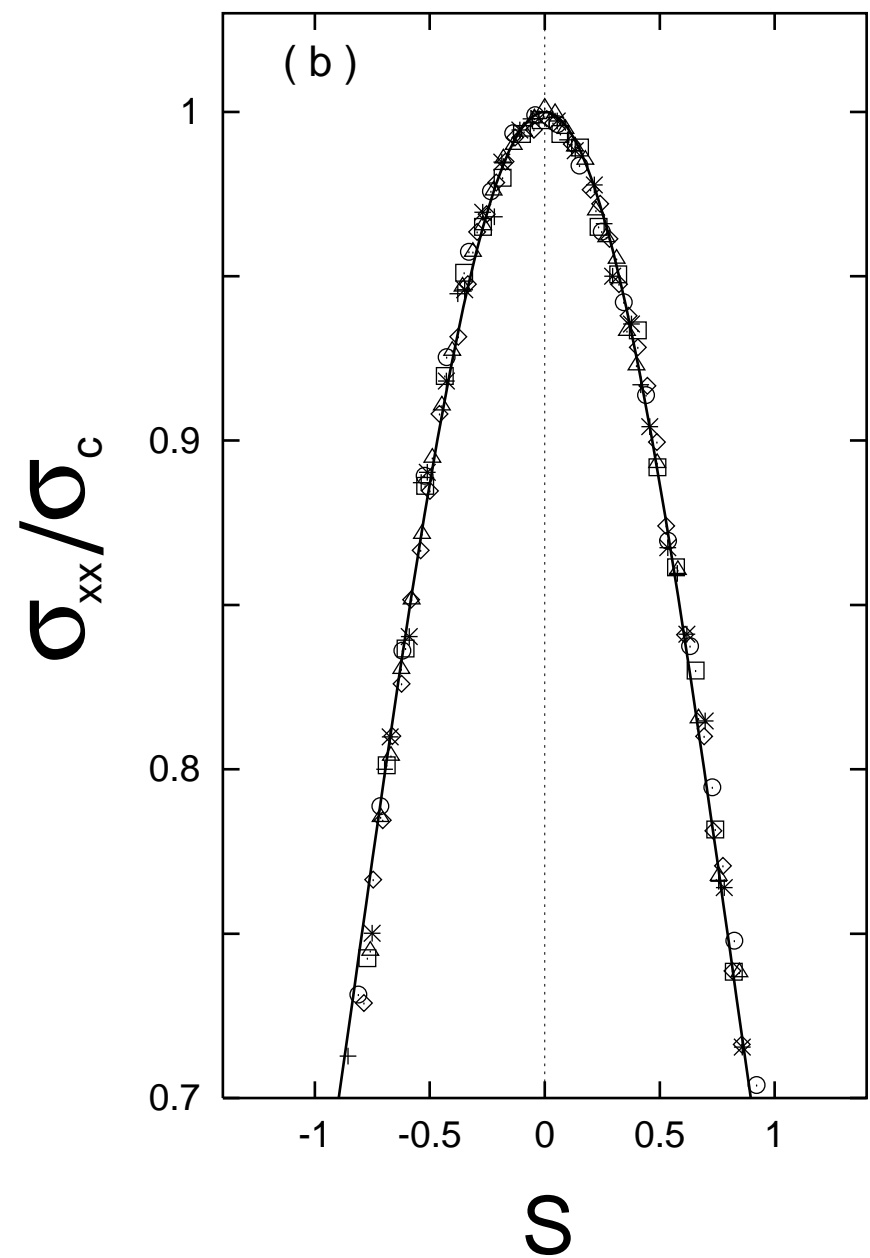
Fig. 2. (a) The single-peak behavior of the longitudinal conductance 
$\sigma_{xx}$ versus $B$, corresponding to the lowest six plateaus to the 
insulator transitions at fixed $n_L=1,2,...,6$, respectively. (b) The 
normalized conductance $\sigma_{xx}/\sigma_c$ ($\sigma_c$ is the peak 
value) versus a scaling variable $s=c_{\nu}(L)(B-B_c)/B_c$ which follows a
universal function form (1).

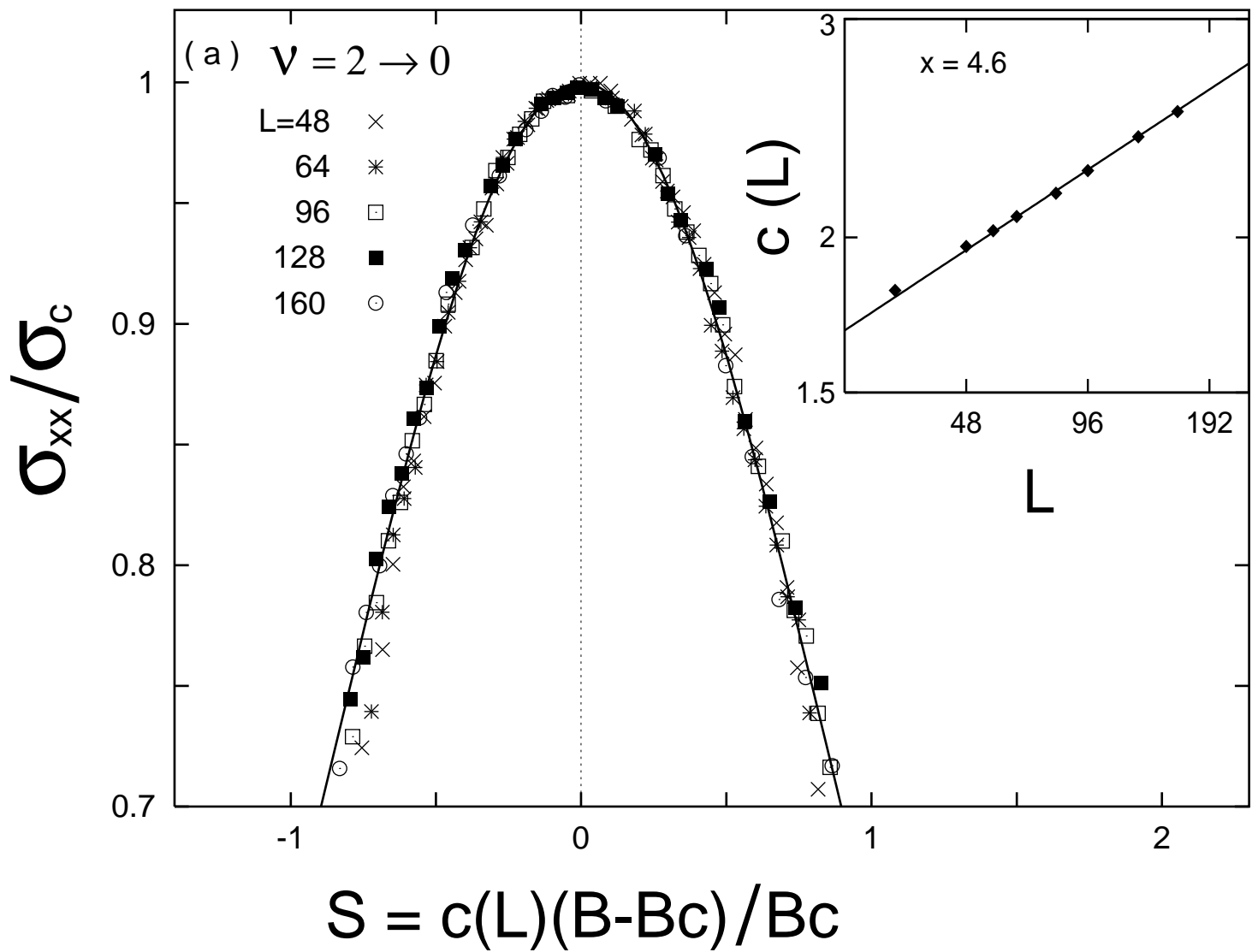
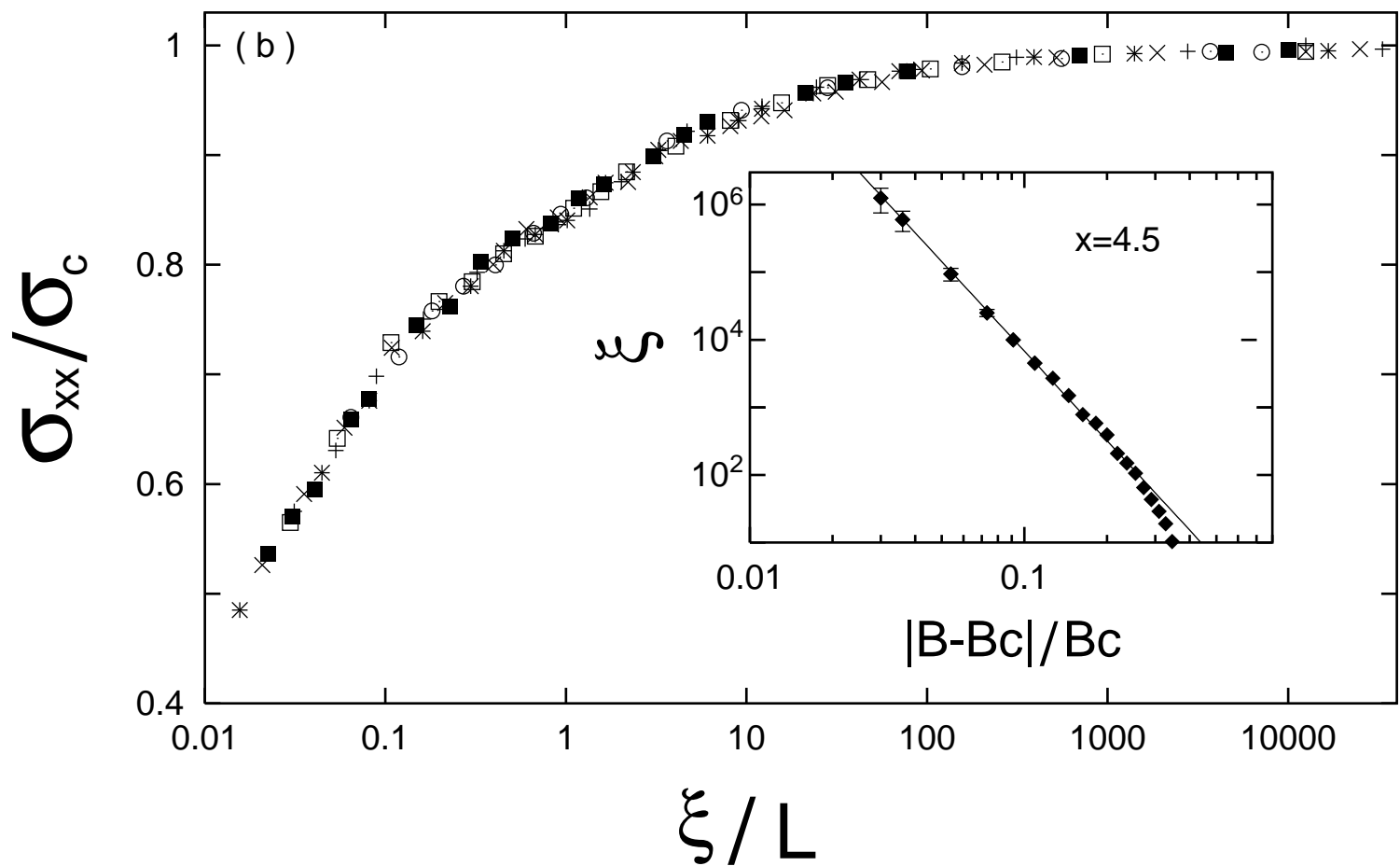
Fig. 3. (a) The normalized conductance $\sigma_{xx}(L)/\sigma_c$ 
for $\nu=2\rightarrow 0$ transition at different sample sizes. The inset: 
$c(L)=c_0L^{1/x}$ where $x=4.6\pm 0.5$. (b) The same data are collapsed  as 
a function of $\xi/L$. The inset: $\xi \propto |B-B_c|^{-x}$.

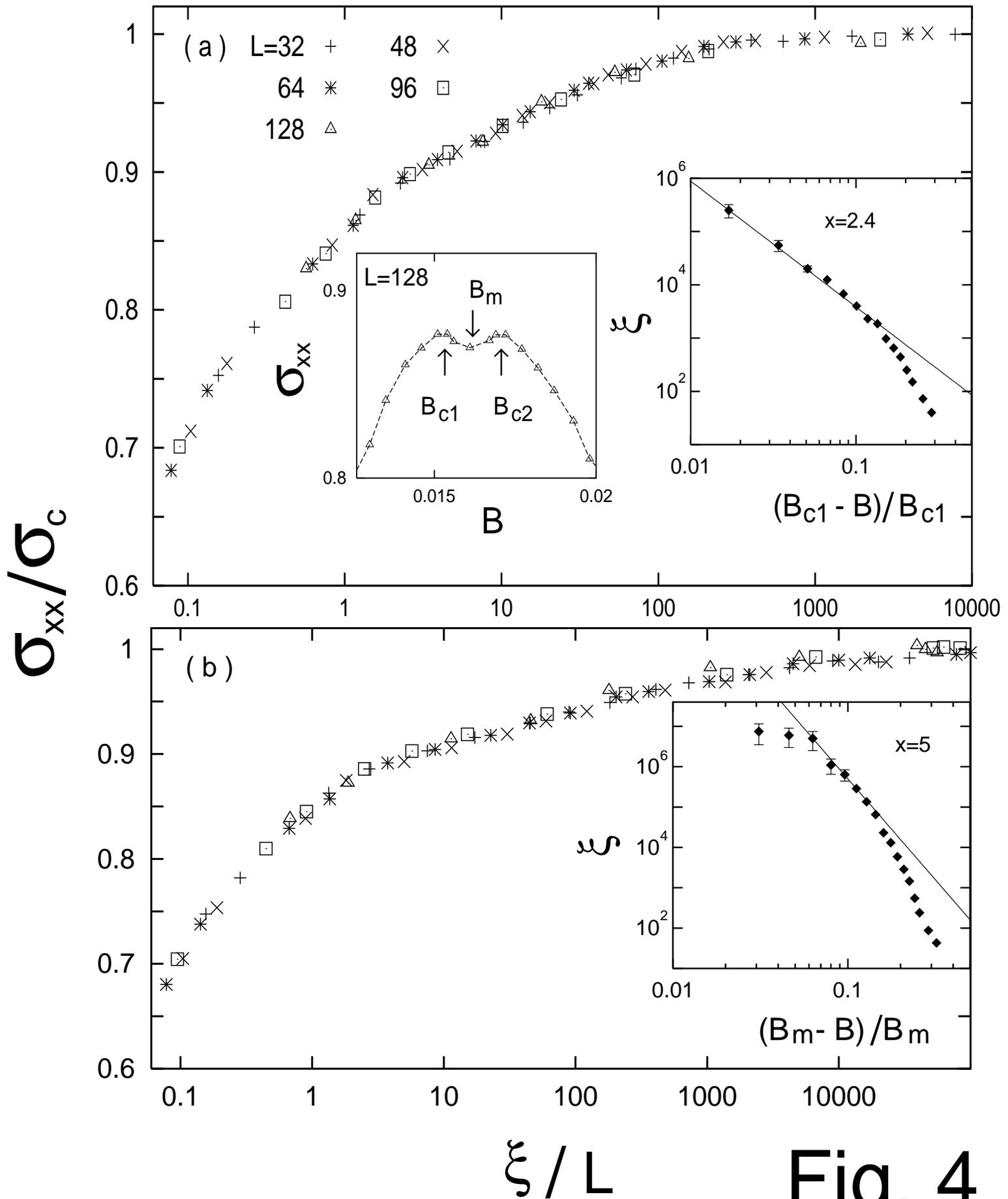
Fig. 4. The double-peak $\sigma_{xx}$ before two lowest extended 
levels merge is shown in the middle panel of (a). (a) 
$\sigma_{xx}/\sigma_c$ as a function of $\xi/L$ by collapsing all the 
data at $B<B_{c1}$ with  $\xi$ shown in the right inset. (b) By assuming 
a single critical point at $B_m$, the data collapsing shows worse quality 
and $\xi$ in the inset becomes saturated approaching $B_m$.

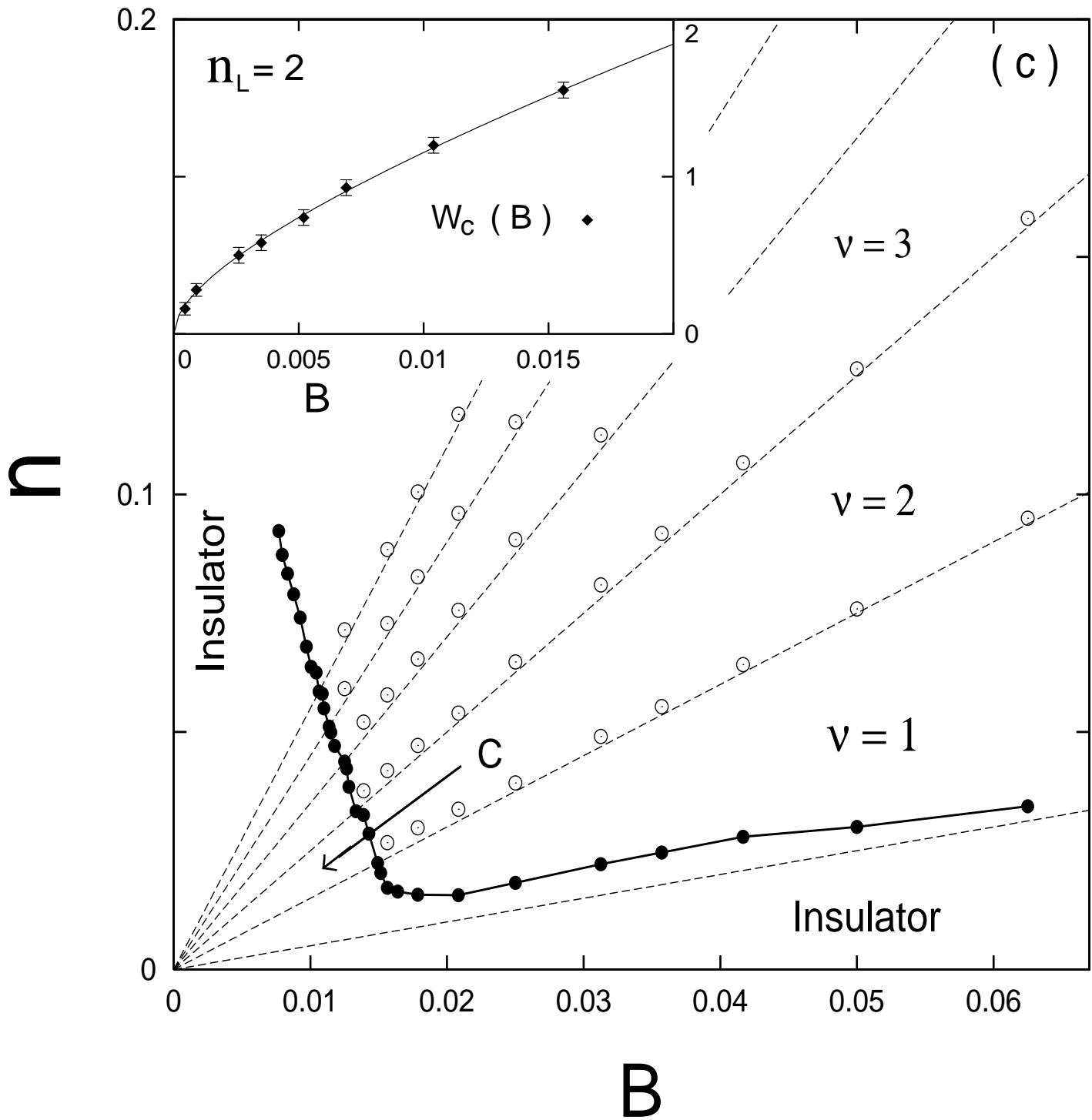
\end{document}